\documentclass[12pt,preprint]{aastex}
\usepackage{psfig}
\usepackage{epsfig}

\newcommand{\cha}{{\sl Chandra}}

\newcommand{\degree}{{\rm o}}

\newcommand{\msun}{M$_{\odot}$}

\newcommand{\xmm}{{\sl XMM-Newton}}
\newcommand{\etal}{{et al.}}

\newcommand{\ergl}{erg~s$^{-1}$}

\newcommand{\GA}{\mbox{\raisebox{-0.6ex}{$\stackrel{\textstyle>}{\sim}$}}}
\newcommand{\LA}{\mbox{\raisebox{-0.6ex}{$\stackrel{\textstyle<}{\sim}$}}}

\slugcomment{Submitted to The Astrophysical Journal.}

\begin{document}

\title{On the Nature of the Bright Short-Period X-ray Source \\
   in the Circinus Galaxy Field}

\author{
  Martin C. Weisskopf\altaffilmark{1},
  Kinwah Wu\altaffilmark{2},
  Allyn F. Tennant\altaffilmark{1},\\
   Douglas A. Swartz\altaffilmark{3},
   and Kajal K. Ghosh\altaffilmark{3}}

\altaffiltext{1}{Space Science Department, NASA Marshall Space Flight Center,
    SD~50, Huntsville, AL 35812, USA}
\altaffiltext{2}{Mullard Space Science Laboratory, University College London,
    Holmbury St Mary, Surrey RH5 6NT, United Kingdom}
\altaffiltext{3}{USRA, NASA Marshall Space Flight Center,
    SD~50, Huntsville, AL 35812, USA}

\begin{abstract}

The spectrum and light curve of the bright X-ray source CG X-1 in the field
of the Circinus galaxy are re-examined.
Previous analyses have concluded that the source is an accreting black
hole of mass \GA\ 50~\msun\ although it was noted that the light curve
resembles that of an AM Her system.
Here we show that the short period and an assumed main sequence companion
constrain the mass of the companion to $<1$~\msun.
Further a possible eclipse seen during one of the \cha\  observations and a
subsequent \xmm\ observation constrains the mass of the
compact object to $<60$~\msun.
If such a system lies in the Circinus galaxy, then the accreting object
must either radiate anisotropically or strongly violate the Eddington
limit.
Even if the emission is beamed, then the companion star which intercepts this
flux during eclipse will be driven out of thermal equilibrium and evaporate
within $\sim10^3$~yr.
We find that the observations cannot rule out an AM Her system in
the Milky Way and that such a system can account for the
variations seen in the light curve.

\end{abstract}


\section{Introduction}
\label{introduction}

\cha\ (see e.g. Weisskopf \etal\ 2002 and references therein) observations have
shown that normal galaxies of the size of the Milky Way (e.g.\ M81, Tennant
\etal\ 2001; Swartz \etal\ 2003) host hundreds of discrete X-ray sources with
luminosities above $10^{36}$~erg~s$^{-1}$.
These sources are comprised of disparate groups of objects, including
supernova remnants, X-ray binaries, and more exotic objects such as
the supersoft X-ray sources.
While the majority of the X-ray emitters are binaries containing a
neutron star or stellar-mass black hole accreting material from a
companion star, some appear to be quite luminous, with implied accretion
rates well above the Eddington limit for a 1.5-\msun\ compact star.
These peculiar bright sources are often referred to as ultra-luminous X-ray
sources (ULXs).

Several explanations for ULXs have been proposed.
They may be intermediate-mass black holes (IMBHs) with masses $\sim 10^2 -
10^4$~\msun, less than the $>10^6$~\msun\ inferred for active galactic nuclei
(Colbert \& Mushotzky 1999; Matsumoto \etal\ 2001; also Ebisuzaki \etal\ 2001).
They may be stellar-mass black holes (SMBHs) in high-mass X-ray binaries with
mildly beamed radiation during a thermal-timescale mass transfer phase (King
\etal\ 2001), perhaps with contributions from synchrotron and inverse-Compton
emission from jets in addition to the thermal X-ray emission from the accretion
disks (K\"{o}rding, Falcke \& Markoff 2002).
They may be long-lasting outbursts in low-mass X-ray binaries, such as in
microquasars (King 2002).
The nature of ULXs is still under debate, in part, because the X-ray
observations alone have difficulty in distinguishing amongst the various
alternatives.

A ULX candidate in the Circinus galaxy ($\alpha_{2000} = 14^{\rm
h}\, 13^{\rm m}\, 12.^{\rm s}3\ $ and $\delta_{2000} = -65^{\degree}\,
20\arcmin\, 13\farcs$, hereafter CG~X-1) appears to be periodic with a period
of $27\pm 0.7$~ks (Bauer \etal\ 2001, Bianchi \etal\ 2002).
Furthermore, at times there is a sharp deep dip in the X-ray light curve
reminiscent of eclipses (Weisskopf 2002, and shown in Figure~\ref{lc_365_356}).
These signatures are important and are useful to determine the nature the
system.
Bauer \etal\ (2001) noted the light curve of CG~X-1 resembles that of AM
Her-type cataclysmic variable (CV) 
system (for a review of AM Her binaries and CVs, see Cropper 1990; Warner 1995),
but
argued against an AM Her interpretation as follows:
(1) No optical counterpart brighter than 25.3 mag is found in {\sl
HST} images, implying that if the source is an AM Her system its
distance is greater than 1.2~kpc, therefore, the X-ray luminosity
is too bright for a typical AM Her system.
(2) The X-ray spectrum does not have the two components (blackbody
and optically thin thermal plasma) typical of AM Her systems.
(3) The inferred period of 27~ks (7.5~hr) is significantly larger than those
of many AM Her systems, which are typically 1.5-4~hr.
(4) The time variability exhibited by CG~X-1 shows flickering similar
to those seen in the Galactic black-hole X-ray sources.
(5) There is only a small probability ($< 0.06\%$) of there being
a foreground star in the field and an even smaller probability that
such a foreground star would be an AM Her system.

Here we re-examine the spectral and timing data of CG~X-1 using data from 
the \cha\ (\S~\ref{chandra}), \xmm\ (\S~\ref{xmm}), and HST (\S~\ref{s:opt})
archives.
We show that the short period and an assumed main sequence companion
constrain the mass of the companion to $<1$~\msun\ (\S~\ref{parameter1}).
Further a possible eclipse  constrains the
compact object to $\LA 60$~\msun\ (\S~\ref{parameter2}).
It is therefore unlikely that CG~X-1 is an IMBH (\S~\ref{parameter2})
or a SMBH in Circinus (\S~\ref{parameter3}).
In \S~\ref{cygX3} we consider the possibility of a He-burning
companion.
Finally, we show that the original arguments of Bauer \etal\ (2001) do  not
conclusively rule out a CV system (\S~\ref{parameter4}).
Although establishing that a short period ULX is present in the Circinus galaxy
would be exciting, we feel that the explanation that this source is a CV in our
own galaxy is also compelling and cannot be dismissed.

\section{The \cha\ Data}
\label{chandra}

The Circinus galaxy has been observed 6 times with the instruments aboard \cha\
on 4 separate dates (Table~\ref{observations}).
The observations were made either with the back-illuminated CCD, ACIS-S3, or
with the High Energy Transmission Grating (HETG)-ACIS-S combination known as the
High Energy Transmission Grating Spectrometer (HETGS).
The spectral fits discussed below and in  \S \ref{xmm} use the default
abundances and cross-sections in the routine PHABS in XSPEC (Arnaud 1996) in
calculating the absorbing column.

 \begin{deluxetable}{lcclll}
  \tablecolumns{6}
  \tablewidth{0pt}
  \tablecaption{{\sl Chandra} Observations of the Circinus Galaxy
\label{observations}}
  \tablehead{
\colhead{ObsID} & \colhead{Date} & \colhead{Exposure} & \colhead{Instrument}
 & \colhead{Frame time} & \colhead{Mode} \\
\colhead{} & \colhead{} & \colhead{ks} & \colhead{} & \colhead{s} & \colhead{} }
  \startdata
355   &  2000 Jan 16  & 1.8 & ACIS-S3    & 0.40/3.20 &  TE$^a$/alternating \\
365   &  2000 Mar 14  & 5.5 & ACIS-S3    & 0.40      &  TE/subframe    \\
356   &  2000 Mar 14  & 23.4& ACIS-S3    & 3.20      &  TE             \\
374   &  2000 Jun 15  & 7.3 & HETGS      & 3.20      &  TE             \\
62877 &  2000 Jun 16  & 61.4& HETGS      & 3.20      &  TE             \\
2454  &  2001 May 02  & 4.5 & ACIS-S3    & 3.20      &  TE             \\
  \enddata
\tablecomments{\\
$^a$ TE --- Timed Exposure
}
 \end{deluxetable}

\subsection{ObsIDs 365 \& 356 \label{subs_365_356}}

Results from the 2000 Mar 14 observation were reported by Smith \& Wilson
(2001), who labeled CG X-1 as source J.
The viewing time of this ACIS-S3 observation was broken into two segments with
different CCD frame times to aid in dealing with pileup.
Initially, a 5.5-ks observation (ObsID~365) was made in a subframe mode with a
frame time of 0.4~s~frame$^{-1}$.
Shortly thereafter, the instrument mode was changed (ObsID~356) to operate in
the full frame mode (3.2~s~frame$^{-1}$).
The time series, uncorrected for pileup, is shown in Figure~\ref{lc_365_356}.
The apparent change in the counting rate in the interval between the two
observations is entirely due to pileup.

In the first data set (ObsID 365), the counting rate was 0.32 counts~s$^{-1}$
(0.15 frame$^{-1}$), and the impact of pileup is small.
In the second data set (ObsID 356), the same 0.32 counts~s$^{-1}$
corresponds to $\sim$1 event~frame$^{-1}$.
Using the simple pileup model given in the \cha\ Proposer's
Guide\footnote[4]{see http://asc.harvard.edu/}, the pileup fraction
is estimated to be $\sim$40\%.
This degree of pileup should lead to an apparent counting rate of 0.19
counts~s$^{-1}$, close to what is observed.

Flickering is detectable during this observation.
Unfortunately, pileup tends to smooth the data at high counting rates
which masks any underlying source variability (as well as
statistical fluctuations).
To test for variability we assumed a steady source and applied a $\chi^2$ test.
We found $\chi^2_\nu=2.51$ for the first segment of 39 data points taken using
the shorter frame time and thus less susceptible to pileup.
We conclude that the source is variable (``flickering''), on time scales of
longer than a single bin in the first segment.
Applying the same test to the last 39 bins of data in the second segment we
found  $\chi^2_\nu=1.20$.
Since the increased amount of pileup would reduce the apparent
variability we conclude that the flickering was more than likely present
throughout the entire observation.

The most obvious feature in Figure~\ref{lc_365_356} is the eclipse-like minimum
seen near the center of the light curve.
During the minimum, the count rate is very low
($\sim2\times10^{-2}~$s$^{-1}$) and appears to be constant, except for
a short flare that extends from 39150 to
39400~s (time as measured in Figure~\ref{lc_365_356}).
If the short flare signals the end of eclipse, then the eclipse duration is
$\sim1500 \pm100$~s.
Examination of these data at higher time resolution suggests that the counting
rate is rising during the flare consistent with what one expects for
a source emerging from eclipse.

Spectral fitting the March data is complicated by the pileup.
The best approach for dealing with pileup in ACIS data (Davis 2001) does not
account for source variability but assumes that the underlying source is
steady.
Table~\ref{fits} lists the results of our fitting various spectral
models to the data that were least impacted by pileup, that from ObsID
365.
All errors in the Table (and in  \S~\ref{xmm}) are 90-percent-confidence single
parameter
uncertainties.
The power law fits were performed with and without corrections for pileup.
Ignoring pileup, we found results consistent with the previous analysis
of Smith \& Wilson (2001).
We applied the Davis (2001) pileup correction as implemented in XSPEC 11.2 with
the pileup parameter $\alpha$ fixed at 0.5 (for small amounts of pileup
allowing $\alpha$ to vary does not improve the fits but simply increases the
uncertainty).
Comparing the two power law fits in Table~\ref{fits} we see that the
pileup-corrected spectral index is steeper as one would expect.

The other spectral fits listed in Table~\ref{fits} also have the pileup
correction applied.
The 10-keV thermal bremsstrahlung model formally provides the best fit amongst
the models.
However, given the uncertainties introduced by applying the correction for
pileup, the underlying variability of the source, and the small variation in the
values of the C-statistic, the other models are equally acceptable.
Therefore, we cannot constrain the nature of the source from these spectral
analyses.

 \begin{deluxetable}{lcccc}
  \tablecolumns{5}
  \tablewidth{0pt}
  \tablecaption{Spectral Fits to ObsID 365 \label{fits}}
  \tablehead{
\colhead{Model} & \colhead{$n_{\rm H}$} & \colhead{Parameter$^a$}
 & \colhead{C-statistic}  & \colhead{Flux$^b$}  \\
\colhead{} & \colhead{$10^{22}$ cm$^{-2}$} & \colhead{}
 & \colhead{646 dof} & \colhead{$10^{-12}$ erg cm$^{-2}$ s$^{-1}$}  }
  \startdata
power law (uncorrected) & 1.02$\pm$0.12  & 1.44$\pm$0.14           & 580.3 & 5.0
\\
power law               & 1.09$\pm$0.12  & 1.60$\pm$0.15           & 580.9 & 5.2
\\
bremsstrahlung          & 1.01$\pm$0.10  & 12.2$_{-3.8}^{+8.7}$    & 578.5 & 5.1
\\
diskbb                  & 0.82$\pm$0.08  & 2.04$_{-0.22}^{+0.27}$  & 579.0 & 5.0
\\
  \enddata
\tablecomments{\\
$^a$Power law index, $\Gamma$;
  or bremsstrahlung temperature, $kT_{\rm br}$, in keV;\\
  or disk blackbody model innermost disk temperature, $kT_{\rm in}$, in keV. \\
$^b$Observed flux (0.5~-~8.0 keV).
               }
 \end{deluxetable}

During the interval covering the $\sim$1500~s at eclipse minimum, 26 counts
were collected.
By \cha\ standards the source is bright.
The best fitting absorbed power law requires a steeper index and a smaller
column than that listed in Table~\ref{fits}.
We also fit these data with a mekal model assuming a column density fixed at
$N_{\rm H} = 1.0 \times 10^{22}$ cm$^{-2}$.
The best fit temperature is $1.5 \pm 0.5$~keV, and the observed flux (0.5 to
8~keV ) is $1.0\times 10^{-13}$ erg~cm$^{-2}$~s$^{-1} (3.1 \times 10^{-13}$
unabsorbed).

For the flare-like event at the end of the eclipse, 19 counts were
collected, and only two had energies less than 2~keV.
If we assume the power law index has not changed from the value listed in
Table~\ref{fits}, the absorbing column would be near $7 \times 10^{22}$
cm$^{-2}$. Such a high column is consistent with what one expects from
eclipsing material in the line of sight.
A much flatter power law with a much smaller column also provides an acceptable
fit (though more difficult to provide an astrophysically sound explanation).

\subsection{ObsIDs 374 \& 62877\label{subs_374_6287}}

The longest consecutive set of \cha\ observations (ObsID's 374 \& 62877)
took place in 2000 June.
The HETGS data were analyzed by Bauer \etal\ (2001) who discovered a $27\pm
0.7$~ks period using the $\sim 1100$ counts from the zero-order image to form
the light curve.
We have repeated this analysis including the first-order flux from
both the High Energy Grating (HEG) and Medium Energy Grating (MEG)
which more than doubled the number of counts to $\sim2500$.
This time series is shown in Figure~\ref{lc_374_62877}.

The 27~ks periodicity is apparent and has been confirmed with long Beppo-Sax
observations (Bianchi \etal\ 2002) and by us ( \S~\ref{xmm}).
The source is ``high'' for about half the period and there are suggestions of
flickering.
Bauer \etal\ (2001) noted that this light curve looks like  one characteristic
of an AM Her type magnetic CV  (see, for example, the light curve of WW Hor,
Tennant \etal\ 1994).
In an AM Her, the matter is being funneled by a magnetic field 
  onto a pole of a white dwarf.
If the polar cap is visible, then the source is bright, 
  but if the polar cap is behind the dwarf, 
  then the source is dim.
This very naturally leads to a square wave profile.
As some matter can fall onto the second pole, 
  it is possible to see some flux in the dim orbital phase.
In this scenario, flickering may arise 
  from inhomogeneities in the accretion flow 
  caused by Rayleigh-Taylor or Kelvin-Helmholtz instabilities
  (Hameury, King \& Lasota 1986) 
  or from thermal instabilities of shocks in the accretion column 
  (Imamura, Wolff \& Durisen 1984; Saxton \etal\ 1998).
Magnetic CVs can also possess an accretion disk, 
  in which case they are referred to as intermediate polars (IPs). 
In the Chandra observations, CG X-1 is dim 
  (less than 10\% of the high rate) 
  for about 25\% of the 27~ks cycle.
If an accretion disc is present,  some of the X-rays emitted from the accreting
pole can be scattered by the accretion disk. 
However, this scattered radiation contributes  no more than 10\% of the flux, at
least during this particular observation.

Analysis of the zeroth order spectrum was reported in Bauer \etal\ (2001).
We extracted the dispersed spectrum with the express purpose to search for line
features.
No strong lines were detected, and we set an upper limit to any narrow
line of about $3.5 \times 10^{-6}$ photons~cm$^{-2}$~s$^{-1}$.
This upper limit corresponds to an equivalent width of 70~eV for a line near
1.7~keV and 450~eV for a line near 6.4~keV.
The underlying continuum emission was well fit by either a thermal
bremsstrahlung model with a temperature of $5.0_{-1.4}^{+2.7}$ keV or a Mekal
model with similar temperature and an abundance of 0.1 of the solar value.

After accounting for pileup, disparate detector configurations, and
the presence/absence of the grating it is apparent that the source
has undergone a change of state between the 2000 March and June
observations.
The average (peak) flux decreased by a factor of 6 (4) and the fraction of time
spent at low counting rates increased from $\sim5$\% to $\sim50$\%.
Because of the uncertainty in the orbital period it is impossible to predict,
with sufficient accuracy, the eclipse location. 
Furthermore, an eclipse during the low rate interval would not
be detectable above the background. 

\subsection{ObsIDs 355 \& 2454}

During the other two \cha\ observations (ObsIDs 355 \& 2454) the flux was low
and highly variable resembling that shown in Figure~\ref{lc_374_62877} in
the time interval (1.08 $-$ 1.17)$\times 10^5$~s, and we were unable to glean
any information from these observations that had not already been discussed by 
Bianchi \etal\ (2002).

\section{The XMM-Newton Observations}
\label{xmm}
{\it XMM-Newton} observed the Circinus galaxy on 2001 Aug 6 for 104 ks (ObsId
0111240101).
We first added together data from the EPIC-PN, MOS1 and MOS2 detectors. 
The summed image is shown in Figure \ref{xmm1} and CG X-1 lies in the wings of
the point spread function (PSF) from the nucleus of the galaxy. 
In addition, another source CG X-2 (using the notation of Bauer et al. 2001) is
also nearby. 
We therefore selected the background region as shown in Figure \ref{xmm1} to
remove possible contamination by both the nucleus and CG X-2.

The background subtracted light curve is shown in Figure \ref{xmm2}. 
The  $\approx$ 27-ks period is obvious. 
Due to a longer observation and the higher number of counts than in the Chandra
data we obtain a better estimate of the period. 
By maximizing the  $\chi^2$ statistic through epoch folding we find a period of
$26250 \pm 150$~s (one sigma error).

The folded light curve is shown in Figure \ref{xmm3}. 
It does not necessarily follow that any one of the dips in the light curve is an
eclipse. However, if we {\it assume} that there is an eclipse, then the dip
centered at phase 0.63 is the best candidate. 
In the Chandra data the count rate during the eclipse was about 5\% of the count
rate outside the eclipse. 
In the {\it XMM-Newton} data  the count rate is below this value for the phase
interval at $0.61 - 0.65$ giving one confidence that the eclipse identification
is reasonable.
The measured eclipse duration depends on the assumed binary period.
Examining this dip for the range of periods consistent with the uncertainty in
our period measurement leads to an eclipse duration of $1000 \pm 100$~s.

The {\it XMM-Newton}-PN spectrum was extracted using the source and background
regions as shown in Figure \ref{xmm1}.
An Fe line was present and was allowed for in our spectral modeling.
Fitting these data to a blackbody did not yield an acceptable fit, with 
$\chi^2/\nu = 970.6/698$. 
A powerlaw provided a better fit, with $\chi^2/\nu = 778.0/696$. 
However, the best-fit model invoked thermal bremsstrahlung ($\chi^2/\nu =
709.8/696$). The parameters of this model are  $N_H=1.24 \pm 0.06 \times
10^{22}$~cm$^{-2}$ and $kT=7.4 \pm 0.6$~keV, consistent with the \cha\
modeling discussed previously (Table~\ref{fits} \& \S \ref{subs_365_356}).
The Fe line had central energy of $6.41 \pm 0.14$~keV,  a width of 
$0.21_{-0.12}^{+0.21}$~keV, and an equivalent width of $230 \pm 57$~eV.  
The detection of the line feature is below the upper limit set by our analysis
of the \cha\  data and thus consistent with that result. 
It is worth noting  that the nucleus of the galaxy has very strong narrow lines
at both 6.4 and 7.0 keV (Sambruna et al. 2001). 
These two lines are also seen in the CG X-1 spectrum  prior to background
subtraction. 
After subtraction the 7.0 keV feature is consistent with zero indicating that
the bulk of the spectral contamination of the nucleus has been properly
removed.
Also, the {\it narrow} 6.4 keV feature also disappeared. 
The spectrum then shows a broad feature consistent with mixture of 6.4 and 6.7
keV lines, which we attribute to intrinsic emission from CG X-1.  
However, we cannot totally rule out that some of the 6.4 keV flux is the
residual from incomplete removal of the contamination  of the galactic nucleus. 
The average unabsorbed flux in the $0.5-8$ keV band was  $1.5 \times
10^{-12}$~erg~cm$^{-2}$s$^{-1}$.

\section{The Optical Data}
\label{s:opt}

We analyzed the two HST WFPC2 images from the observations made
on 1996 Aug 11 with the F606W filter.
The two images were combined to remove cosmic rays.
The coordinates were adjusted so that the
nucleus would match in both the \cha\ and HST data sets.
The region near CG X-1 is shown in Figure~\ref{f:hst}.

The circle in Figure~\ref{f:hst} is centered at the \cha\ position
of CG X-1 and has a radius of 0.5 arcsec.
There is some weak diffuse emission inside the circle and the location of a
point-like source automatically detected with the analysis software is
indicated.
This source is only 0.25 arcsec from the \cha\ position and within
our ability to register the two data sets.
The HST magnitude in the F606W filter for the point source is 24.3.
If we include all flux inside a circle of radius 0.23 arcsec (5 HST pixels)
minus the flux in the annulus from 0.23 to 0.36 arcsec, then $m_{\rm F606W}$
becomes 23.5.
This emission can be seen in both original images and thus is unlikely
to be an artifact of the cosmic ray removal process.
There is a suggestion of this source in Figure 2 of Bauer \etal\ (2001) where
one sees a slight excess at the bottom of the innermost contour.
Thus, a conservative upper limit to an optical counterpart is
$m_{\rm F606W}$=24.3 and it is quite possible that there is, in
fact, an optical point source near the \cha\ source.

\section{The Nature of the X-ray source}

\subsection{Constraints implied by the orbital period}
\label{parameter1}

Here, we adopt the interpretation that the 26.2~ks (7.3~hr) period that we infer
from the \xmm\ data is the orbital period of a binary.
We further suppose that it is a semi-detached binary system (as it is an X-ray
source) and the orbit is circular.
In this case, the separation of the centers of the two stars, $a$, is given by
\begin{equation}
  \biggl({ a\over {{\rm R}_\odot} }\biggr)\ =\ 2.35~
  \biggl( {{M_1+M_2}\over {{\rm M}_\odot}} \biggr)^{1/3}
  \biggl({P\over{10~{\rm hr}} }\biggr)^{2/3} \ ,
\end{equation}
and the secondary star's effective Roche-lobe radius, $R_{\rm h}$, is given by
\begin{equation}
   {R_{\rm h}\over a} \
      = \ {{0.49q^{2/3}} \over {0.6q^{2/3}+\ln(1+q^{1/3})}}
\end{equation}
(Eggleton 1983), where $P$ as the orbital period, $M_1$ and $M_2$ are the masses
of the primary and secondary stars respectively, and $q\ (=M_2/M_1)$ is the
mass ratio.
It follows, for a large range of $q$, that the mean density of a Roche-lobe
filling secondary star is
\begin{equation}
  {\bar \rho}_2 \ \approx 0.78~{\bar \rho}_\odot
    \biggl({P \over {10~{\rm hr}}}\biggr)^{-2}  \ ,
\label{meanden}
\end{equation}
where $4\pi {\bar \rho}_2 R_2^3 /3 = M_2$, and $R_2$ is the radius of the
secondary.
Thus, knowing the orbital period allows one to determine the mass-radius
relation of the secondary star.

We show in Figure~\ref{mass_radius} the Roche-lobe radius of the secondary
as a function of its mass for the 26.2 ks orbital period.
In the same figure we show also the mass-radius relations of a star
at the zero-age main-sequence (ZAMS) and at the stage that it begins
to evolve away from the main sequence (TAMS).  (The case of a core of a more
massive evolved star is considered in \S~\ref{cygX3}.)
From Figure~\ref{mass_radius} we can see that for a variety of masses of the
compact star --- from that typical of white dwarfs (0.7 and 1.0~M$_\odot$) to
stellar-mass black holes (10~M$_\odot$) to the small version of IMBHs
(50~M$_\odot$) --- that the near-main sequence mass-donor (secondary) star
has a mass \LA 1~M$_\odot$.

\subsection{Constraints implied by the eclipse duration}
\label{parameter2}

X-rays are emitted from near the surface of the compact star in binary systems
and so the emission region is much smaller than the orbital separation.
If the binary has a sufficiently large orbital inclination $i$, then the
emission region can easily be eclipsed by the mass-donor star.
The eclipse fraction is
\begin{equation}
  \Delta_{\rm ec} \ =\ {1 \over \pi}~
   \cos^{-1} \left( \ {1 \over {\sin i}}
  \sqrt{1-\bigg({R_2\over a }\bigg)^2} \right)
\label{eclipse_duration}
\end{equation}
(eqn. 2.64 of Warner 1995), and its maximum value is
\begin{equation}
  {\rm max} \{ \Delta_{\rm ec} \} \ = \
     \frac{1}{\pi}\ \cos^{-1} \sqrt{1 - \biggl( \frac{R_2}{a}\biggr)^2} \ ,
\end{equation}
where $\Delta_{\rm ec} = \delta t_{\rm ec}/P$ and $\delta t_{\rm ec}$ is the
eclipse duration.

We use the observed $\Delta_{\rm ec}$ and Equation (4) to constrain the size of
the mass-donor star and the orbital inclination.
Furthermore, we can set constraints on the mass ratio, $q$, through the
transcendental equation
\begin{equation}
  \ln (1+q^{1/3}) \ =\ q^{2/3}~
    \biggl( {0.49 \over { \sin \pi \Delta_{\rm ec}}} -0.6 \biggr) \ .
\end{equation}

We show in Figure~\ref{duration} the eclipse fraction, $\Delta_{\rm ec}$, as a
function of the inverse of the mass ratio, $q^{-1}(=M_1/M_2)$ for different
orbital inclinations, $i$.
In \S~\ref{subs_365_356} we found that the eclipse duration was 1500~s.
In \S~\ref{xmm} we found the eclipse duration to be 1000~s. 
An eclipse duration of 1000~s corresponds to $\Delta_{\rm ec} = 0.038$,
implying that the maximum value for $q^{-1}$ is 60 (which requires
an extreme orbital inclination of $90^{\circ}$).
Since the secondary (mass-donor) is less massive than $1.0$~M$_\odot$
(\S~\ref{parameter1}), this now implies that the mass of the compact star is
$\LA  60$~\msun.
For modest orbital inclinations around 70$^{\circ}$, both the \cha\  and \xmm\
data
suggest that the compact star is much less massive than 10~\msun.

In summary, an orbital period of 7.3~hr implies a mass \LA 1~\msun\ for a main
sequence secondary star.
The measured eclipse duration (1000~s $-$ 1500~s) requires the $M_1/M_2$ ratio
to be
smaller than 60.
Thus, if CG~X-1 is an eclipsing binary, it certainly can not contain an IMBH
with a mass $10^2-10^4$~\msun.

\subsection{CG X-1 as a Stellar Mass Black Hole in Circinus}
\label{parameter3}

If CG~X-1 is a binary belonging to the Circinus galaxy and it contains
a SMBH of mass $\sim$10\msun\ with a main-sequence companion, then its X-ray
emission must be very anisotropic  or, otherwise, greatly violate the Eddington
limit.
Assuming isotropic emission, the observed X-ray flux corresponds to $\sim
10^{40}$~\ergl\ at the 4 Mpc distance of Circinus.
To avoid exceeding the Eddington limit, the emission must be beamed into a cone
confined to about 10\% of the sky.
Assuming an eclipsing system, a substantial fraction of the beam must
therefore illuminate the companion star.
We estimate that, if this is true, the energy intercepted by the secondary star
in CG X-1 in one year is $\sim 10^{46}$~erg.

We have deduced from the orbital period (\S~\ref{parameter1}) that
the companion star has a mass $<1.0$~\msun.
Main-sequence stars of this mass have an intrinsic luminosity $\sim
10^{33}$~erg~s$^{-1}$, several orders of magnitude less than the power in
X-rays that the secondary star in CG X-1 intercepts.
The gravitational binding energy of a star of mass $M$ and radius $R$ is
\begin{equation}
   E_{\rm b} \sim \frac{G M^2}{R} \ .
\end{equation}
For a star of mass $\sim$1.0~\msun,
  $ E_{\rm b}  \sim 4 \times 10^{48}$~erg.
If X-ray emission from CG X-1 is beamed, the large amount of energy
intercepted by the secondary star not only drives the star out of
thermal equilibrium but also can evaporate it within a mere $\sim
10^3$~yr.
Since it is very unlikely that we have happened to come upon such
a short-lived system, we argue that CG~X-1 is probably not a SMBH
in the Circinus galaxy.

\subsection{CG X-1 with a core He-burning companion}
\label{cygX3}

Our arguments against the interpretation that CG X-1
is a IMBH or a SMBH are based on the two assumptions:
(1) it is an eclipsing system; and (2) the companion
is a main-sequence or a slightly evolved star.
The assumption of a main-sequence or a slightly evolved star is
reasonable, considering stars spend the vast majority of their lifetimes
at this evolutionary phase.
However, if the black-hole interpretation is retained, the companion star could
be at a more evolved stage or have undergone a common envelope phase.
In close binary systems, the compact remnant may spiral into the
envelope of a massive companion, removing the hydrogen-rich layers,
and leave an exposed core (perhaps with a tenuous, low-mass, envelope)
mass-donor star.
The orbital period of CG~X-1 can only constrain the mean density of the
companion star, thus a more massive compact core within the
Roche-lobe radius can be accommodated.

An example of a system that may have evolved in this manner is Cyg~X-3
(e.g., van~den~Heuvel \& de~Loore 1973).
The orbital period of 4.8~hr and the spectral similarity of the secondary
to a WN8 Wolf-Rayet star (Koch-Miramond \etal\ 2002) are consistent with
a core He-burning companion for Cyg~X-3 of about 2~\msun.
However, at the distance of the Circinus galaxy, the companion would
have to be very bright in the visible (\S~\ref{s:opt}).
If we assume the companion is of low mass (say 2~\msun) then this implies that
the optical flux is fainter than the upper limit we set in \S \ref{s:opt} and
that the optical identification is incorrect.
But then, the lifetime problem discussed in the previous section is again
relevant.

Thus, while we cannot definitively rule out the possibility of a He-core
companion, there are arguments against this scenario.
This possibility should be re-examined once the properties of X-ray sources with
He-burning companions are better established.

\subsection{CG X-1 as a foreground AM Her system}
\label{parameter4}

The orbital parameters that we have derived are consistent with those
of AM Her systems (see e.g.\ Downes \etal\ 2001).
We now comment on the arguments against the AM Her binary interpretation
put forth by Bauer \etal\ (2001) and summarized in \S\ref{introduction}.

(1) The lack of a bright optical counterpart:  One can always make a star
appear fainter by moving it further away.
Thus there are two parts to this argument.
First the companion must be located within our Galaxy, and second
the inferred X-ray luminosity cannot exceed the effective Eddington
limit for a CV.
The Galactic coordinates of the Circinus galaxy are $l$$=$311.3, $b$$=$-3.8.
This places Circinus near the Galactic plane and along a line-of-sight that
passes the closest to the Galactic center at a distance of 5 kpc from the Sun.
Thus, a reasonable distance estimate for purposes of discussion would be 5 kpc.
There is a tendency for longer period AM Her systems (Patterson 1984)
to have earlier spectral types (K5).
If we assume a K5-V spectral type (M$_V$=7.3, V$-$R=1.0), the distance
of 5 kpc and 5 magnitudes of visual extinction inferred from the X-ray
absorbing column, one would have m$_V$=25.8 and m$_R$=23.8.
The HST F606W filter bandpass includes flux from both V and R bands.
If we assume the rate is dominated by the R-band flux then a distance
slightly greater than 5 kpc would account for the HST data.
On the other hand, if the spectral type is later than K5, the inferred distance
would be less.
We note that the long period AM Her system V1309~Ori (=RX~J051542+01047;
Garnavich \etal\ 1994) has an M0 companion.
Thus there is no difficulty accounting for the optical flux.\footnote[5]{If CG
X-1 is a magnetic CV and if its optical luminosity is dominated by emission
from the accreting material instead of from the companion star, then the HST
brightness upper limit implies that the system could easily be outside the Milky
Way, thus contradicting the AM Her interpretation. 
However, an XMM survey of AM Her systems finds that there is a roughly equal
chance that an AM Her system is in an X-ray-on and in an X-ray-off state
(Ramsay, G., private communication). 
In the X-ray-off state, the optical flux cannot be from an accretion disk as
there is no accretion taking place. 
Thus it is possible that the optical emission of the source is dominated by the
emission of the companion star during the HST observation. 
Further simultaneous X-ray and deep optical observations can clarify this
issue.}     

In the above we have used the X-ray column to infer the reddening.
This is reasonable since both the X-ray column and interstellar
reddening are due to higher Z material (metals) and that it is possible that the
interstellar medium has been enriched along this line of sight which is close
to the galactic plane.  
Further, we note that no other source in the Circinus galaxy has an X-ray column
significantly smaller than that seen for CG X-1 (Bauer \etal\ (2001)) and thus
the argument that the column is not due to local extinction is bolstered. 
Conversely, one can assume that the Dickey \& Lockman (1990) radio measurement
of neutral hydrogen tells us the Galactic X-ray column, in which case the
reddening should be based on a column of $6\times10^{21}$ cm$^{-2}$. 
For a typical gas to dust ratio, and assuming any
intrinsic column does not contribute to the reddening, we now would have only 3
magnitudes of extinction.  
For this case, the assumed K5V star would need to be 2.5 times further away
placing it 12.5 kpc from the Earth or 9.4 kpc from the Galactic center, still
within our galaxy.  
Of course, the inferred X-ray luminosity would be 6 times greater which would
make CG X-1 the most luminous accreting mCV ever observed.
Even under these conditions, the case for an mCV is not ruled out. 

The second part of the argument has to do with the inferred X-ray luminosity.
Using the brightest observed flux and an assumed distance of 5 kpc
gives a luminosity of $2\times 10^{34}$~erg~s$^{-1}$ (0.5$-$8 keV).
For an accreting white dwarf the corresponding Eddington limit is
$f \times L_{\rm ed}$ where $f$ is the fraction of the white dwarf
covered by the polar cap and $L_{\rm ed}$ is the Eddington limit for
spherical accretion.
For the inferred luminosity and a 0.5 M$_\odot$ white dwarf, we
find $f \GA\  2 \times 10^{-4}$ which is not unreasonable
(see Wu \& Wickramasinghe 1990).
Although this hard X-ray luminosity is far higher than the average
bolometric of AM Her-type systems ($2\times 10^{32}$ erg/s Ramsay \&
Cropper 2003), it has been observed in another accreting magnetic
cataclysmic variable.
The unusual system GK Per, a very long period Intermediate Polar,
has been observed at $1.6 \times 10^{34}$ ~erg~s$^{-1}$ in the hard 2-20
keV Ginga band (Ishida \etal\ 1992).

To rule out an AM Her system one would have to show that most reasonable
values for spectral type, reddening, assumed distance, etc. are not allowed.
Since even the extreme case considered here is allowed, we conclude that an AM
Her cannot be ruled out.

(2) The lack of a soft X-ray spectral component:
Some AM Her systems exhibit both a soft X-ray emission pole and a hard X-ray
emission pole as does the prototype system AM Her itself (Heise \etal\ 1985).
The emission from the hard pole is mainly bremsstrahlung of $\sim 10$~keV, and
the emission from the soft pole is black-body radiation at a temperature $\sim
10-100$~eV.
The high column ($\sim 1 \times$ 10$^{22}$ cm$^{-2}$) associated with CG X-1
makes it essentially impossible that such a low temperature black-body could
have been
detected.
Further, some AM Her systems only show the hard X-ray pole (e.g.\ CE Gru,
Ramsay \& Cropper 2002).

(3) The orbital period exceeds that of most AM~Her systems:
The 7.3-hr period of CG~X-1 is, in fact, shorter than the 7.98-hr period of the
AM Her system V1309~Ori (Garnavich \etal\ 1994).
Provided that the white dwarf has a magnetic moment large enough to lock the
binary into synchronous rotation (see e.g.\ Wu \& Wickramasinghe 1993), the
orbital period of an AM Her system does not have to be in the  range between
1.5 and 4~hr.

(4) The source shows quasi-periodic like variability:
AM Her binaries also show strong X-ray flickering, 
e.g.\ AM Her (Heise \etal\ 1985) and UZ For (Ramsay \etal\ 1993).

(5) The probability of a foreground AM~Her system in the field is low:
This is perhaps the strongest argument against the AM~Her  interpretation. 
The Circinus galaxy lies close to the Galactic Plane ($b=-3.8$), which
increases the chance of detecting a foreground star in this field.
In fact, Bauer \etal\ (2001) calculated that they expected one
background source in the 2 arcmin radius from the Circinus
galaxy nucleus where they found 16 sources.
The Log-N - Log-S curve of the Circinus galaxy steepens above 
$1 \times 10^{-13}\ {\rm erg~cm}^{-2}{\rm s}^{-1}$ 
($2\times 10^{38}\ {\rm erg~s}^{-1}$) 
and the Log-N - Log-S curve for the Galactic Plane is fairly flat in this region
(index $=-0.8$, Sugizaka \etal\ 2001).
Thus, in looking toward the Circinus Galaxy, the brighter sources are more
likely to be in our Galaxy.
Nevertheless, the chance probability for such a source in the field, and further
for that source being a magnetic CV is low. 
However, the argument that CG~X-1 is not a foreground object in the
Milky Way is statistical, and if only one object is considered there
can always be an exception --- however small the probability.

\section{Summary}

\begin{figure}
\begin{center}
\epsfig{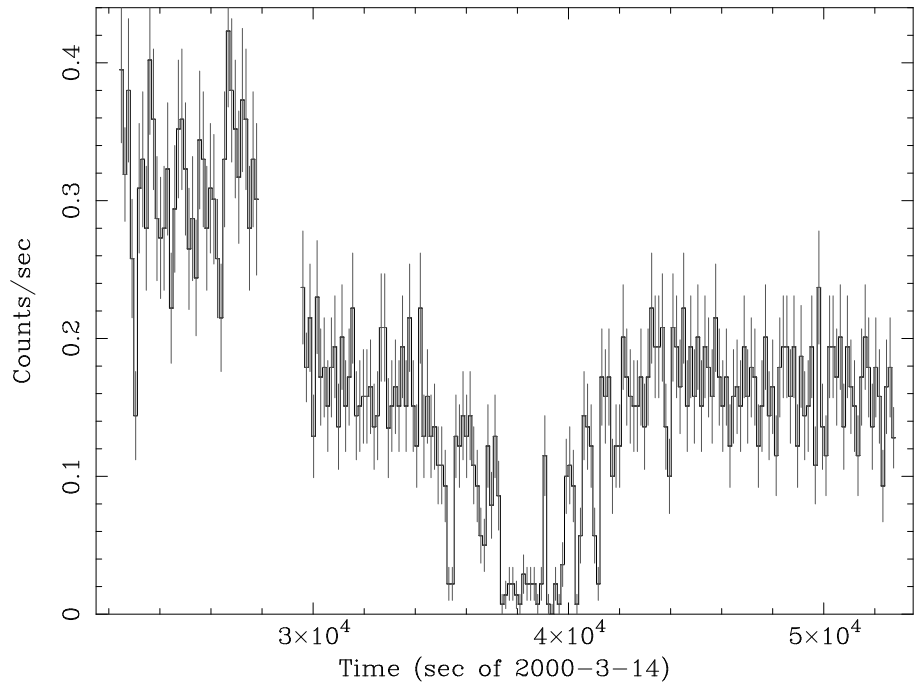}
\end{center}
\figcaption{
Counting rate versus time, in bins of 140~s, for the data from CG~X-1
obtained in March of 2000 (ObsIDs 365 and 356).
The apparent change in the count rate near 28,000 s is due to increased
pileup accompanying the change in the frame time from 0.4 s to 3.2 s.
\label{lc_365_356}}
\end{figure}

\begin{figure}
\begin{center}
\epsfig{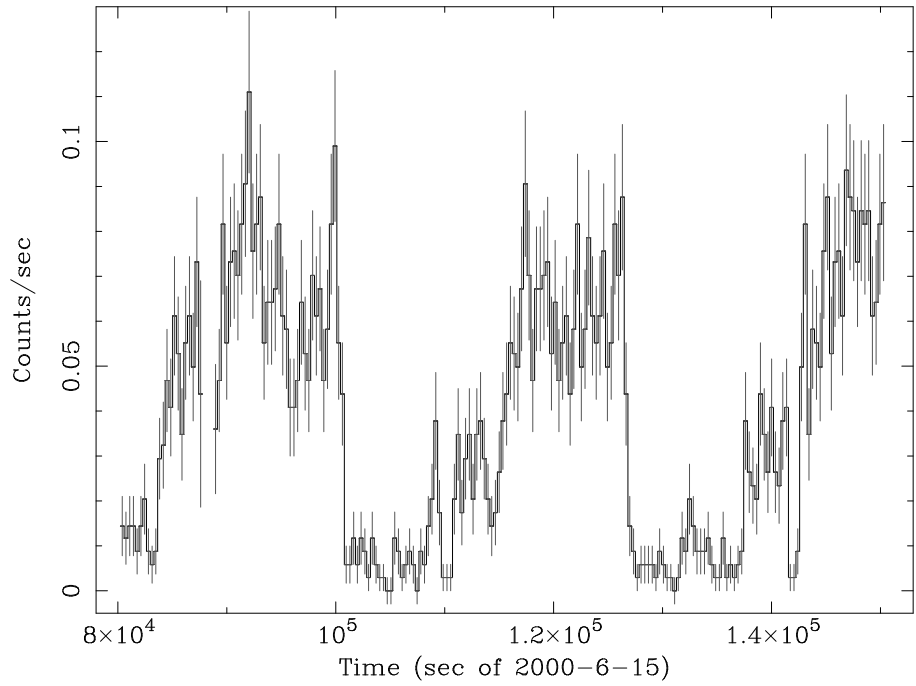}
\end{center}
\figcaption{
Counting rate versus time, in bins of 342~s, for the data from CG~X-1
obtained in June of 2000 (ObsIDs 374 and 62877).
\label{lc_374_62877}}
\end{figure}

\begin{figure}
\begin{center}
\epsfig{file=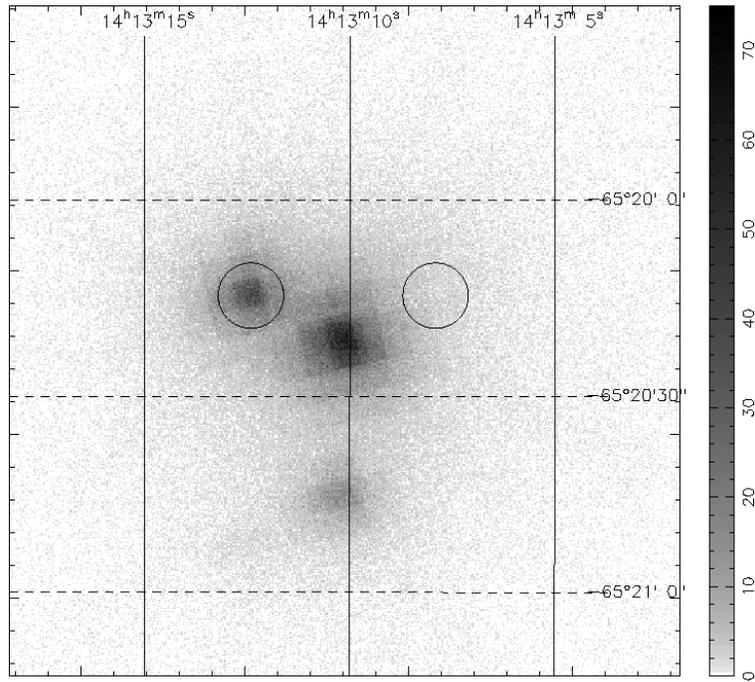,width=9cm,angle=-90}
\end{center}
\figcaption{The combined {\it XMM-Newton} PN/MOS1/MOS2 image of  the Circinus
galaxy.  
The brightest source at the center is the nucleus.  
The circle above and left of the nucleus is the CG X-1 source region.  
The source below the nucleus is  CG X-2.  
The circle above and to the right of the nucleus is the region used for
estimating  the background.
\label{xmm1}}
\end{figure}

\begin{figure}
\begin{center}
\epsfig{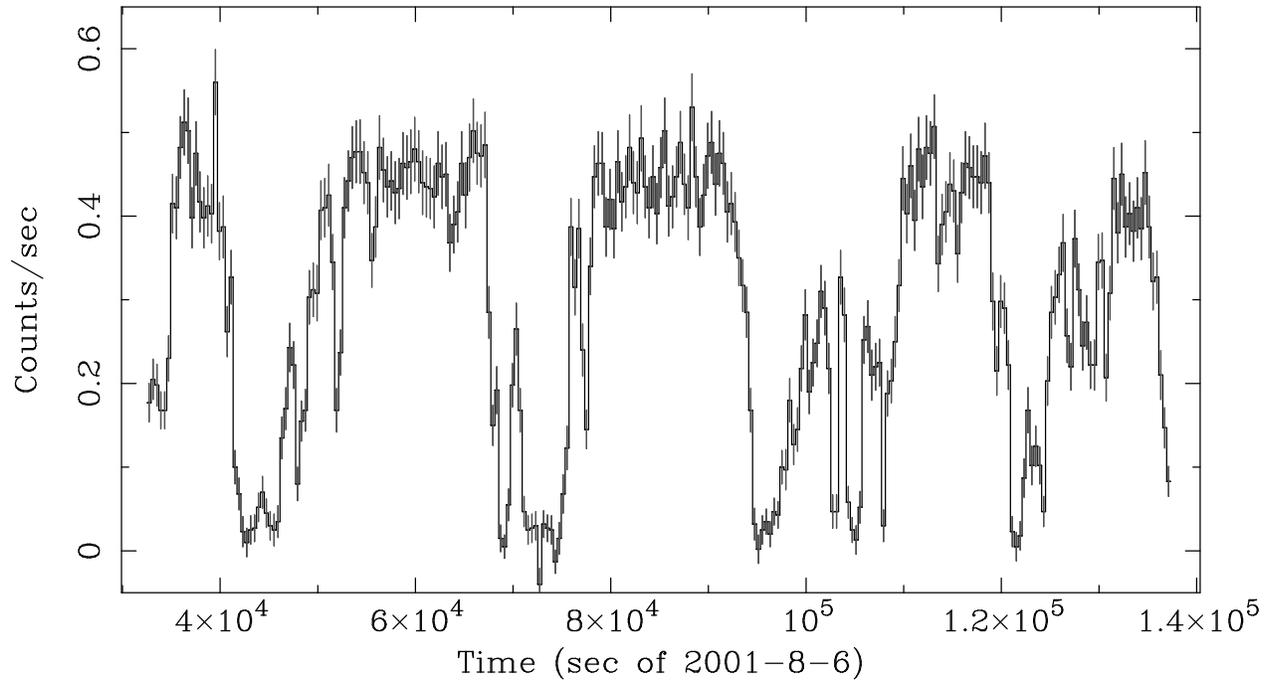}
\end{center}
\figcaption{The PN/MOS1/MOS2 summed light curve from CG X-1 in 400-s bins.
\label{xmm2}}
\end{figure}

\begin{figure}
\begin{center}
\epsfig{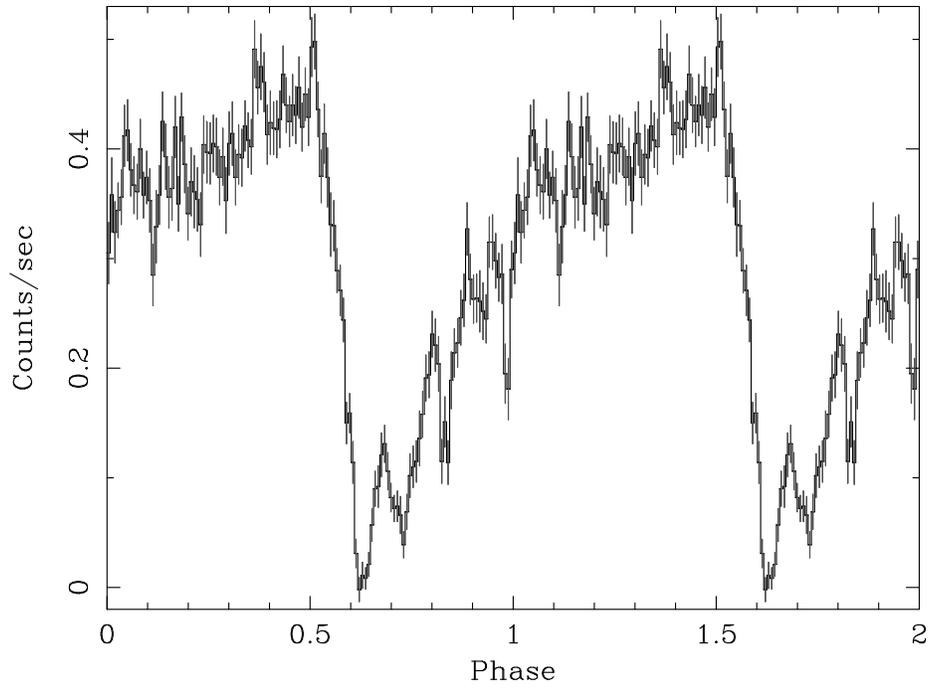}
\end{center}
\figcaption{
The PN/MOS1/MOS2 folded light curve from CG X-1 using a period of 26250 s.
\label{xmm3}}
\end{figure}

\begin{figure}
\begin{center}
\epsfig{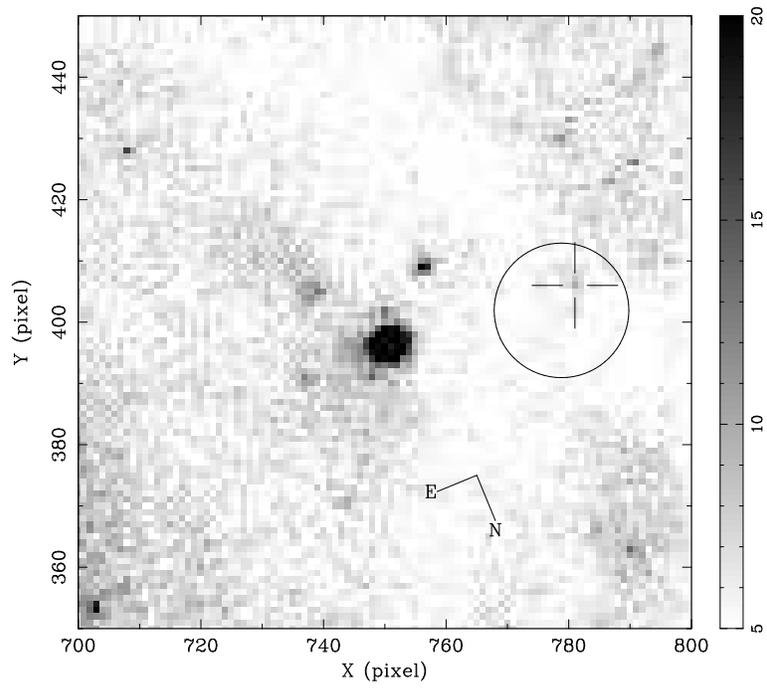}
\end{center}
\figcaption{
HST WFPC2 F606W image of the region including CG X-1.
The circle encompasses a region $0\farcs5$ in radius about the X-ray position of
CG X-1.
A weak optical source located within this region is indicated.
\label{f:hst}}
\end{figure}

\begin{figure}
\begin{center}
\epsfig{file=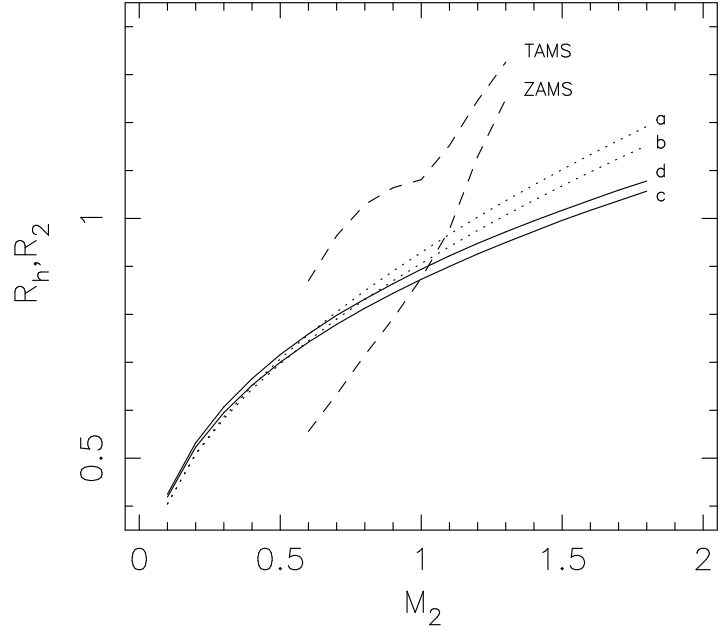,width=10cm}
\end{center}
\figcaption{
  The Roche-lobe radius, $R_{\rm h}$, of the secondary star in a binary
with an orbital period of 26.2~ksec is shown as a function of the secondary
star mass, $M_2$.
Curves a and b (dotted lines) are for a mass of the primary of 0.7 and of
1.0~M$_\odot$ respectively; curves c and d (solid lines) are for a primary
mass  of 10 and of 50~M$_\odot$ respectively.
The two dashed curves are the mass-radius relations (M$_{2}$ - R$_{2}$) of stars
in the zero-age-main-sequence (ZAMS) stage and of stars beginning to evolve
toward the giant stage -- terminal-age-main-sequence (TAMS).
The stellar radius, $R_2$, is derived from the evolutionary tracks
calculated by Bressan \etal\ (1993).
The ZAMS stage is assumed to correspond to point 1 in the track of Bressen et
al. (1993) and the slightly evolved stage to point 2 with
the metalicity $Y=0.28$ and $Z=0.02$.
$R_{\rm h}$, $R_2$, and $M_2$ are in solar units.
\label{mass_radius}}
\end{figure}

\begin{figure}
\begin{center}
\epsfig{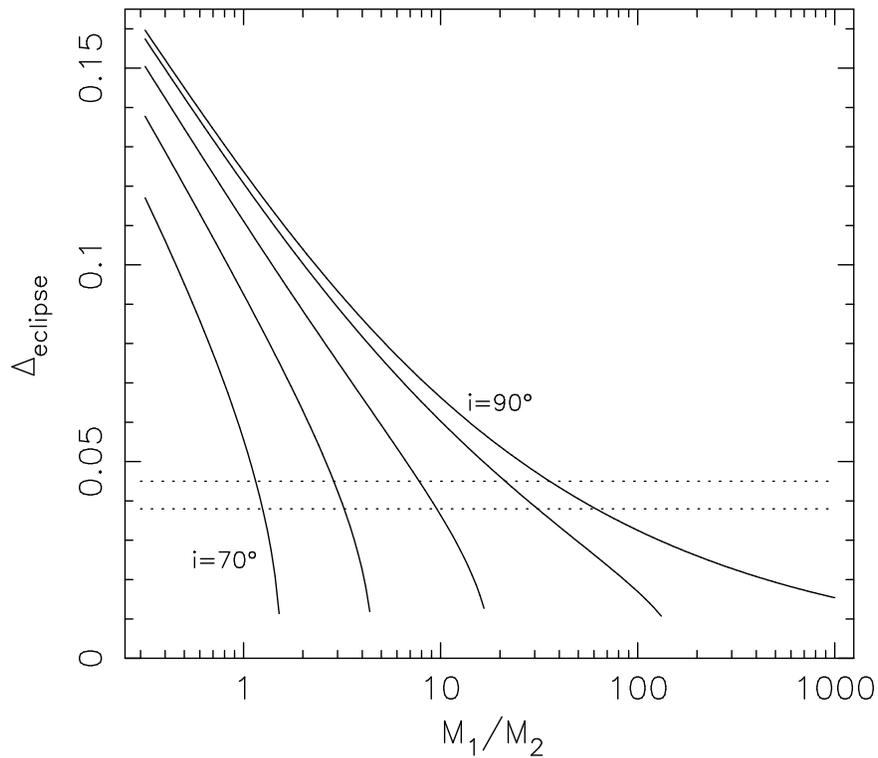}
\end{center}
\figcaption{
The eclipse fraction, $\Delta_{\rm ec}$ (=$\delta t_{\rm ec}/P$), is shown as a
function of the mass ratio $M_1/M_2$ (=$q^{-1}$).
The solid lines from left to right are for orbital inclinations $i$ =
70$^\circ$, 75$^\circ$, 80$^\circ$, 85$^\circ$ and 90$^\circ$.
The dotted lines represent the observed bounds on $\Delta_{\rm ec}$ from \xmm\
(lower) and {\cha} .
\label{duration}}
\end{figure}

We presented spectral and timing analysis of the x-ray source CG~X-1 using
archival \cha, \xmm, and HST data.
The \xmm\ spectrum was  best fit by a thermal bremsstrahlung model with  $kT=7.4
\pm 0.6$~keV.
Based on the interpretation that the observed period is orbital and an assumed
main sequence companion, the companion star is \LA 1~\msun.
A possible eclipse constrains the mass of the primary compact star to less than
60~\msun, thus ruling out an IMBH.
Further, we showed that if one assumes the source is a SMBH binary in the
Circinus galaxy, then the high luminosity of the X-ray source coupled with the
small orbital separation and small mass of the companion would quickly drive
the companion out of thermal equilibrium.
The high X-ray luminosity in this case also causes the companion to evaporate
within $10^3$~yrs.
These arguments make it unlikely that CG X-1 is associated with the
Circinus galaxy.
Based on our analysis of extant HST data we have shown that optical
observations do not rule out a K5 or later companion star in the Milky Way.
Finally, we emphasize that the roughly square-wave light curve
(Figure~\ref{lc_374_62877}) is similar to that of AM Her systems where the
accretion is funneled onto a single pole of a white dwarf and an accretion
shock is formed.
When the source counting rate is high throughout the orbit (as in
Figure~\ref{lc_365_356}), accretion is either occurring onto two poles of the
white dwarf or a disk has formed.
Such changes are naturally explained 
by the presence of the magnetic white dwarf 
(see e.g.\ Warner 1995 and references therein).
While the chance probability of finding an AM Her system in this field is small,
the AM Her interpretation is consistent with all the data and cannot be ruled
out. Further X-ray and optical observation are called for to firmly establish
the nature of this source. 

\acknowledgments

KW and DAS thank the hospitality of the Aspen Center for Physics, where
part of this work was carried out. We also wish to thank the anonymous referee
for pointing out the potential for an evolved companion.

\newpage

\end{document}